\documentclass[aps,pre,twocolumn,amssymb,10pt]{revtex4-1}

\usepackage{graphicx}
\usepackage{dcolumn}
\usepackage{bm}
\usepackage{natbib}
\usepackage[utf8x]{inputenc}
\usepackage{physics} 
\usepackage{hyperref}
\usepackage[mathlines]{lineno}
\usepackage{color}
\usepackage{amsthm}
\usepackage{amsmath}
\usepackage{amssymb}
\usepackage{mathrsfs}
\usepackage{array}
\usepackage{graphicx}
\usepackage[normalem]{ulem}
\usepackage{verbatim,bm}
\usepackage[normalem]{ulem}
\usepackage{xfrac}
\usepackage{bbold}
\usepackage{enumerate}
\usepackage{multirow}

\newcommand\Bigger[2][7]{\left#2\rule{0mm}{#1truemm}\right.}

\begin{document}

\preprint{APS/123-QED}

\title{ Universal entanglement entropy in the ground state of biased bipartite systems      }

\author{Ohad Shpielberg$^{1,2}$}
\email{ohads@sci.haifa.ac.il}
\affiliation{$^1$Haifa Research Center for Theoretical Physics and Astrophysics, University of Haifa, Abba Khoushy Ave 199, Haifa 3498838, Israel}
\affiliation{$^2$Department of Mathematics and Physics, University of Haifa at Oranim, Kiryat Tivon 3600600,  Israel}




\begin{abstract}

The ground state entanglement entropy is studied in a many-body bipartite quantum system with either a single or multiple conserved quantities. It is shown that the entanglement entropy exhibits a universal power-law behaviour at large $R$ -- the occupancy ratio between the two subsystems. Single and multiple conserved quantities lead to different power-law exponents, suggesting the entanglement entropy can serve to detect hidden conserved quantities. Moreover, occupancy measurements allow to infer the bipartite entanglement entropy. All the above results are generalized for the R\'enyi entropy.   

\end{abstract} 

\maketitle


Entanglement has been harnessed as a resource in quantum sensing, quantum computing and quantum communication \cite{wilde2013quantum,preskill2018quantum,NielsenChuang}. The promising potential of quantum technologies has led to intense study of entanglement quantification, especially for many-body systems \cite{Entanglement_Review_Horodecki,Plenio_Entanglement_Review,Entanglement_MB_Vedral,Horodecki_PRXQuantum}. Furthermore, methods of detection, manipulation and certification of entanglement  have been proposed, bridging the gap between theory and experiment \cite{guhne2009entanglement,Huber_EntanglementCost_Review,shpielberg2020diffusion,turkeshi2022enhanced,frerot2022unveiling,omran2019generation}. The bulk of above mentioned work suggest that optimally measuring and controlling entanglement is system specific. Moreover,  the optimal method differs according to purpose. This highlights the appeal to find universal properties of entanglement, applicable to a wide range of systems.

One such universal property is displayed by the area law of ground state entanglement entropy \cite{Eisert_AreaLaw}.  For a lattice system of characteristic size $L^d$ governed by a local Hamiltonian, the ground state is expected to scale like $L^{d-1}$ (with possible logarithmic corrections), different from the volume-like entanglement entropy expected for generic states. However, the area law does not provide a good estimate for the ground state entanglement entropy of many-body systems with finite $L$. Indeed, a universal estimate for such ubiquitous systems is lacking.

Here, we focus on finite $L$ systems with a conserved quantity, e.g. $N$ particles (fermions or bosons) occupying a composite system. Biasing the system's particles to mostly occupy subsystem $A$, we study the ground state entanglement entropy. This setup corresponds to: (a) A box partitioned by a finite-sized piston, allowing particle transfer between the subsystems via tunneling. Moving the piston to reduce (expand) the volume of subsystem $B$ ($A$) controls the bias. (b) An interacting many-body system in the presence of an asymmetric double well potential, separating subsystem $A$ (left potential well) from subsystem $B$ (right potential well). The offset between the wells controls the bias. (c) A many-body system with a strong inter-particle attraction in the presence of an asymmetric double well potential. The bias is governed by the strength of the inter-particle attraction and the potential offset explicitly breaks the symmetry  \cite{jaksch1998cold,greiner2002quantum,links2006two,Shpielberg_PowerLaw}.  

Already in \cite{Shpielberg_PowerLaw}, it was demonstrated for a few choice models, both closed and open quantum systems, exhibits a power-law decay of the ground state Von Neumann entanglement entropy.
The purpose of this letter is to prove that at large bias, the ground state entanglement entropy of closed systems exhibits a universal structure. More precisely, for a Hamiltonian system with a single conserved quantity defined by the observable $\hat{N}=\hat{N}_A + \hat{N}_B$ and where $\hat{N}_X$ acts on the subspace $X\in \lbrace A,B \rbrace$, the ground state entanglement entropy with a fixed charge $\langle \hat{N} \rangle = N $ has a universal power-law structure:  $S_{EE} \propto \frac{\ln R}{R}$ at large $R$ values, where $R = \langle \hat{N}_A\rangle / \langle \hat{N}_B\rangle   $ is the bias. The result is independent of the particular setup or the bias driving mechanism \cite{Shpielberg_PowerLaw}. 
     
The universal power-law decay is obtained also when more than one conserved quantity is considered. Here, $R$ is the bias with respect to one chosen conserved quantity (a precise definition will follow). It is later on argued that the entanglement entropy does not necessarily vanish in this case as $R\rightarrow \infty$. Nevertheless,  $\Delta S_{EE}(R \gg 1) \propto R^{-1/2}$ \footnote{An exception is discussed later on. }, where $\Delta S_{EE}(R) \equiv S_{EE}(R)-S_{EE}(\infty) $, implying the decay towards $S_{EE}(\infty)$ maintains a universal power-law structure. 

A universal power-law decay is similarly obtained for the R\'enyi entropy $S_q(R)$, both for the single conserved quantity and for multiple conserved quantities, suggesting experimental accessibility in many body systems \cite{islam2015measuring}. 

From the above statements, it should be understood that the entanglement entropy and R\'enyi entropy not only quantify quantum correlations. They allow to infer the particle number in the composite system $N$ from the particle measurement in the dilute subsystem $A$. and to differentiate between a single conserved quantity to multiple conserved quantities. Uncharacteristically, the entanglement entropy gives direct information about observables of many-body bipartite systems.




Before proving the main results, it is useful to study a physical model, susceptible to analytical and numerical treatment. Let us consider $N$ interacting spinless fermions occupying a $1D$ system of $2L$ lattice sites. The system Hamiltonian is given by  
\begin{eqnarray}
\label{eq:FH fermions}
    \hat{H}_{IF} &=& - t \sum^{2L-1} _{j=1} \hat{c}^\dagger _{j+1} \hat{c}_{j} + \hat{c}^\dagger _{j} \hat{c}_{j+1} + U \sum^{2L-1} _{j=1} \hat{n}_j \hat{n}_{j+1}
       \\ \nonumber
    &&
    + \mu \sum^{2L} _{j=L+1}   \hat{n}_j + \frac{U}{2} (\hat{n}_1 + \hat{n}_{2L})  ,
\end{eqnarray}
where $\hat{c}_j(\hat{c}^\dagger _j)$ are fermionic annihilation (creation) operators at site $j$ with the anti-commutation relation $\lbrace \hat{c}_j , \hat{c}_k \rbrace = \delta _{j,k}$ and the number operators are 
 $\hat{n}_j \equiv  \hat{c}^\dagger _j  \hat{c}_j $.   The nearest neighbors interaction strength is controlled by $U$. 
 Similar to the double well potential case (b), $\mu$ represents the potential offset, partitioning the system into the subsystems $A$ and $B$ with  $j\in 1,...,L$ and  $j=L+1,...,2L$ correspondingly. 
 The generality of the results is unaffected by the boundary terms in \eqref{eq:FH fermions}, which simplify the analytical analysis. 
 One can verify that $ \hat{H}_{IF}$ and $\sum^{2L} _{j=1}  \hat{n}_j $ commute, implying that the system's particle number $N$ is conserved.  

The interacting fermions model \eqref{eq:FH fermions} can be mapped using the Jordan-Wigner transformation \cite{Coleman2015Book,suppmat} onto the $XXZ$-like model in $1D$ of localized   $\sfrac{1}{2}$ spins:  
\begin{eqnarray}
\label{eq:FH spin Hamiltonian}
    \hat{H}_{S}  &=& -t \sum^{2L-1} _{j=1} \left( \hat{S}^+ _j \hat{S}^- _{j+1} +\hat{S}^- _j \hat{S}^+ _{j+1}  \right)   + U \sum^{2L-1} _{j=1} \hat{S}^z _j \hat{S}^z _{j+1}
    \nonumber \\ 
    && +  \mu \sum^{2L} _{j=L+1} 
       \hat{S}^z _j, \textrm{ with } \hat{S}^\pm _j = \frac{1}{2}(\hat{S}^x _j \pm i \hat{S}^y _j).
\end{eqnarray}
The particle number is expressed by  $N = L + \sum^{2L} _{j=1} \langle \hat{S}^z _j \rangle  $, where $N=0,1,...,2L$. Notice that the aforementioned boundary terms of \eqref{eq:FH fermions} are swallowed in \eqref{eq:FH spin Hamiltonian}.

Let us analyze a few analytically tractable cases. First, we set $N=1$ and $L=1,2$. We analyze the system at large bias, i.e. at large $\mu $ values. The two cases provide the necessary intuition to  prove the universality of the entanglement at large $R$ values.

The $\ket{\pm}_j$ states denote the $\hat{S}^z $ eigenstates at site $j$. Namely, $\hat{S}^z _j \ket{\pm}_j = \pm \frac{1}{2} \ket{\pm}_j   $. For $N=1$, any state can be spanned using the orthonormal set 
\begin{equation}
\ket{k} \equiv \ket{-}_1 ... \ket{-}_{k-1} \ket{+}_k \ket{-}_{k+1} ... \ket{-}_{2L} . 
\end{equation}
Notice that   $ \ket{k}  $  are the eigenvectors of $\hat{S}^z _j $  with eigenvalues $ (-\frac{1}{2}+\delta_{j,k}) $ for any $j,k$. In this $N=1$ subspace and up to shifting by a constant, we can write the Hamiltonian in the $\ket{k}$ basis: 
\begin{eqnarray}
\label{eq:Hs 1particle}
\hat{H}_S &=& -t\sum^{2L-1} _{k=1} \ket{k}\bra{k+1} + \ket{k+1}\bra{k} 
\\ \nonumber
&&+ \mu \sum^{2L} _{k=L+1}  \ket{k}\bra{k} + \frac{U}{2}\sum_{k=1,2L}\ket{k}\bra{k}.
\end{eqnarray}

To calculate the entanglement entropy and R\'enyi entropy, it is useful to write the ground state in its Schmidt decomposition. For $1\leq k \leq L$, $\ket{k} = \ket{k}_A \ket{vac}_B$ and for $L+1\leq k \leq 2L $,   $\ket{k}= \ket{vac}_A \ket{k-L}_B$. Here, $\ket{vac}_{X}$ corresponds to eigenstate of $\hat{S}^z _j$  with eigenvalue $-\frac{1}{2}$ for all $j\in X$ where $X=\lbrace A,B\rbrace $ and $\ket{k}_X$ is defined similarly to $\ket{k}$, but restricted to the subsystem $X$. Finally, for the rest of this work, $\lambda = -t/\mu$ and $u = -U/4t$. Focusing on the large $\mu$ limit ($|\lambda|\ll 1$), $R$ is expected to become large. The ground state can be recovered from a standard perturbation theory analysis where $ \hat{H}_S / \mu   =  H_0 + \lambda \hat{V} $ at $|\lambda|\ll 1$.  


\textbf{\textit{L=1:}}
In this simple two-site case, the ground state can be obtained analytically \cite{suppmat}.   
At $\lambda=0$, the ground state is non-degenerate. The perturbative analysis gives the ground state $\ket{G}$ 
\begin{eqnarray}
\label{eq:gs N=1 L=1}
    \ket{G} &=& (1-\frac{1}{2}\lambda^2) \ket{1} - \lambda \ket{2} +  \mathcal{O}(\lambda^3) 
    \\ \nonumber 
    &=&  (1-\frac{1}{2}\lambda^2) \ket{1}_A \ket{vac}_B - \lambda \ket{vac}_A \ket{1}_B +  \mathcal{O}(\lambda^3) .
\end{eqnarray}
From \eqref{eq:gs N=1 L=1}, we obtain $R = 1/\lambda^2 +\mathcal{O}(1/\lambda) $. The leading order inverse quadratic dependence of $R$ on $\lambda$ is rather general as will be argued later on. The Schmidt decomposition implies any pure state can be written as $\ket{\psi} = \sum_i \alpha_i \ket{u_i}_A\ket{v_i}_B$, where  $\ket{u_i},\ket{v_i}$ are orthonormal sets and $\alpha_i$ are the Schmidt coefficients.  The entanglement entropy is given by $S_{EE} = -\sum_i |\alpha_i|^2 \ln |\alpha_i|^2$. Using again \eqref{eq:gs N=1 L=1} and the $R(\lambda)$ relation, we recover to leading order $S_{EE}(R)= \frac{\ln R}{R}$. The R\'enyi entropy is defined as  $S_q = \frac{1}{1-q}\ln\left( \sum_i |\alpha_i|^{2q} \right)$.  For the ground state \eqref{eq:gs N=1 L=1}, we recover  $S_{q}(R)= \frac{1}{1-q} R^{-q}  $ if $0<q<1$,  and  $S_{q}(R)=  \frac{q}{q-1} R^{-1} $ for $q>1$. Both to leading order in $R$.

\begin{figure}
    \centering
    \includegraphics[scale=0.5]{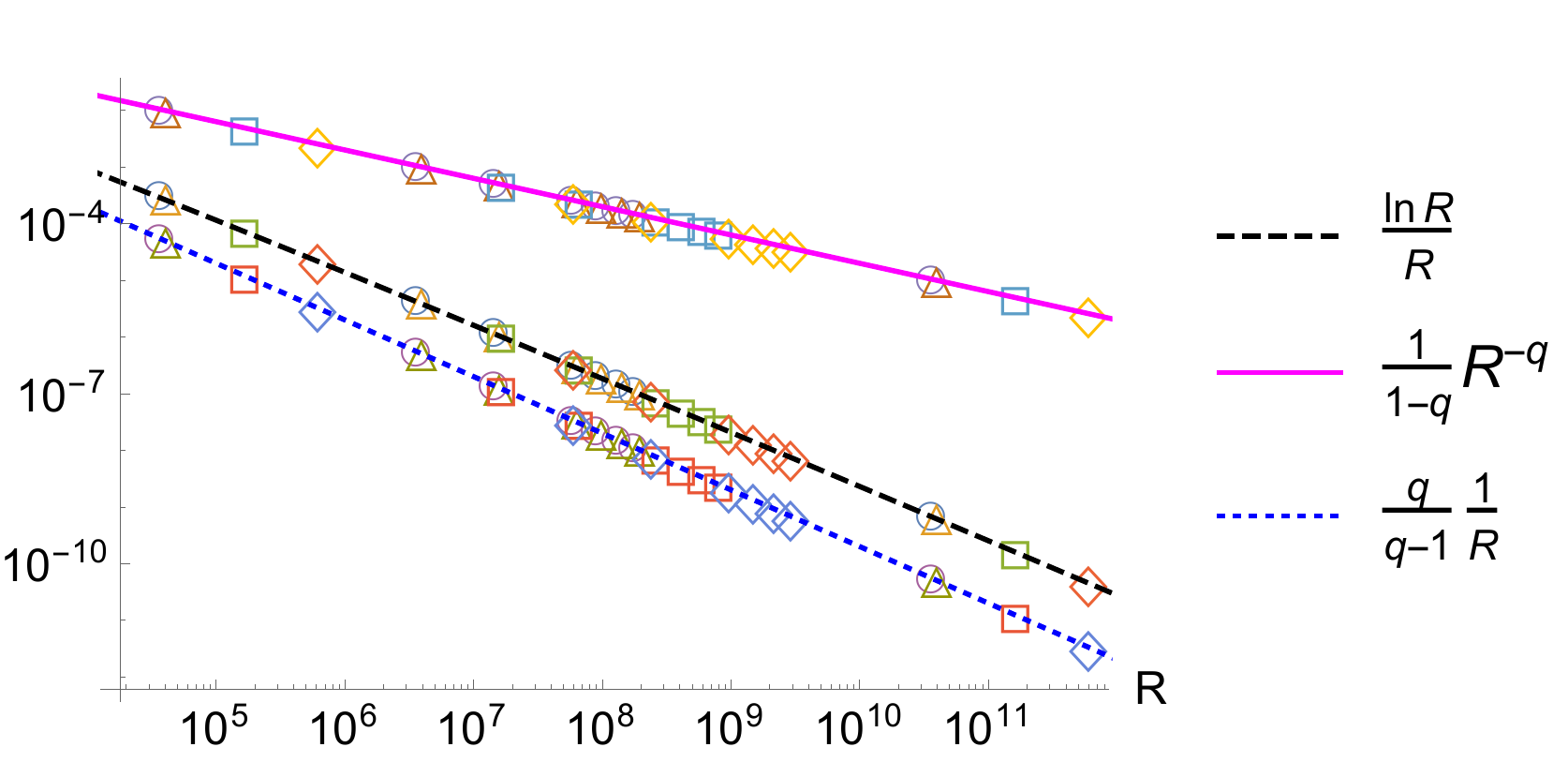}
    \caption{ The entanglement entropy and R\'enyi entropy with $q=\lbrace \frac{1}{2},2 \rbrace$ are numerically calculated in the range   $\mu \in \left[10^2,10^5 \right]$ and for four parameter sets $(U,t,L)$ at $N=1$.
Circles: $(\frac{1}{2},1,3)$; Triangles: $(0,1,3)$; Squares: $(1,1,6)$; Diamonds: $(2,1,10)$. All the points attain large $R$ values and collapse onto the universal curves $S_{EE} = \frac{\ln R}{R}$ (black dashed line), $S_q = \frac{1}{1-q}R^{-q}$ (magenta solid line) for $0<q=\frac{1}{2}<1$ and $S_q = \frac{q}{q-1}\frac{1}{R}$ (blue dotted line) for $1<q=2$. No fitting parameters are required, further validating the analytical predictions.      }
    \label{fig:IF N=1 universal plot}
\end{figure}

\textbf{\textit{L=2:}} The ground state at $\lambda=0$ is degenerate and is spanned by the states  $\ket{1},\ket{2} $. In the particular case of $u=0$, the degenerate perturbation theory implies
\begin{eqnarray}
\label{eq:gs N=1 L=2}
    \ket{G} &=& \frac{1}{\sqrt{2}}\left((1-\frac{\lambda}{4})\ket{1}_A - (1+\frac{\lambda}{4})\ket{2}_A \right) \ket{vac}_B
    \nonumber  \\ 
    &&  -\frac{\lambda  }{\sqrt{2}} \ket{vac}_A \ket{1}_B  + \mathcal{O}(\lambda^2). 
\end{eqnarray}
We find to leading order $R=2/\lambda^2$ and 
\begin{eqnarray}
\label{eq:N1L2 entanglement}
    S_{EE}(R) &=& \frac{\ln R}{R}   \\ \nonumber 
          S_q(R)  &=& 
        \Bigger[10]\{\begin{array}{@{}cl}
                 \frac{1}{1-q}  R^{-q} & \text{if }0<q<1 \\[3mm]
                 \frac{q}{q-1}\frac{1}{R} & \text{if }q>1,
        \end{array}
\end{eqnarray}
The analysis for $u\neq 0$ is carried out in \cite{suppmat}. It is shown there that \eqref{eq:N1L2 entanglement} remains valid for any $u$ value. Numerically, Fig.~\ref{fig:IF N=1 universal plot} demonstrates that when setting  $N=1$, \eqref{eq:N1L2 entanglement} holds for any $L$ and $u$ values and with no fitting parameters.  The $N=2$ case is addressed in \cite{suppmat}, leading to the replacement $R\rightarrow R/N$  in \eqref{eq:N1L2 entanglement}. 

At this point, the lessons of the interacting fermions model \eqref{eq:FH fermions} can be generalized. In what follows we present the proof for the power-law decay of the entanglement 
associated with a single conserved quantity.  Consider an isolated composite system $AB$ of finite dimensional Hilbert space. The Hamiltonian $\hat{H}_{AB} $ has a single conserved quantity associated with the operator $\hat{N} = \hat{N}_A + \hat{N}_B$ where $\hat{N}_X$ acts only on subsystem $X$. Moreover, $\hat{N}_X$ is assumed to have a non-degenerate lower bound, set to zero for simplicity.

Assume a perturbative analysis exists, i.e.  $\hat{H}_{AB} = \hat{H}_0 + \lambda \hat{V}$ for small $|\lambda|$ values and a fixed charge $\langle \hat{N} \rangle = N$ \cite{macieszczak2019coherence}. The ground state can be written as $\ket{G(\lambda)} =\frac{1}{\sqrt{\mathcal{N}_\lambda}} \left( \ket{G^{(0)}} + \lambda \ket{G^{(1)}} + O(\lambda^2) \right)   $, where $\ket{G^{(0)}},\ket{G^{(1)}}$ are orthogonal \footnote{This is always the case for the standard non-degenerate and degenerate perturbation theory.}. Normalizing $\ket{G^{(0)}} $, implies $\mathcal{N} _\lambda = 1 + \braket{G^{(1)}}{G^{(1)}} \lambda^2$ is a normalization constant for $\ket{G(\lambda)}$, truncated at the order of the perturbation theory. Here, the perturbation to first order will be sufficient. At the limit of $\lambda \rightarrow 0$, we assume the the particles are all biased to occupy system $A$. Namely, we can write  $\ket{G^{(0)}} = \ket{\Phi_N}_A\ket{vac}_B$  with ${}_A\langle \Phi_N | \hat{N}_A \ket{\Phi_N}_A = N $ and $ {}_B\langle vac |   \hat{N}_B \ket{vac}_B = 0$. Note that since there is a single conserved quantity, the state $\ket{vac}_B$ is unique. Next, one can write \cite{suppmat}
\begin{equation}
 \ket{G^{(1)}} = \sum^N _{n\geq 0} \alpha_{n,k} \ket{\phi^{k} _{N-n}}_A \ket{\psi^{k} _{n}}_B  
\end{equation}
 as a Schmidt decomposition. $ \lbrace \ket{\phi^{k} _{n}}_A, \ket{\psi^{k} _{n}}_B \rbrace_{n,k}$ are orthonormal eigenstates of the operators $\hat{N}_A , \hat{N}_B$ correspondingly with eigenvalues denoted by the subscript $n$. 
Note that $\ket{\psi^{k} _{n=0}}_B = \ket{vac}_B$. Therefore, the truncated ground state in the Schmidt representation is
\begin{eqnarray}
\label{eq: Schmidt decomp single conserved quantity}
     \ket{G(\lambda)} &=& \frac{1}{\sqrt{\mathcal{N}_\lambda} } \ket{\Phi^\lambda _N} _A \ket{vac}_B 
       \\ \nonumber 
     &&
     +\frac{\lambda}{\sqrt{\mathcal{N}_\lambda} } \sum^N _{n> 0,k} \alpha_{n,k} \ket{\phi^{k} _{N-n}}_A \ket{\psi^{k} _{n}}_B,  \nonumber \\  \nonumber
     \ket{\Phi^\lambda _N}_A &=& \ket{\Phi_N}_A + \lambda \sum_k \alpha_{0,k}  \ket{\phi^k _N}_A.
\end{eqnarray}
It is important to realize that $ \langle \Phi_N \ket{\phi^k _N}=0$ due to the orthogonality of $\ket{G^{(0)}},\ket{G^{(1)}}$. This is the crucial point, separating between the single and multiple conserved quantity analysis.  Using \eqref{eq: Schmidt decomp single conserved quantity}, we find  $R \propto \frac{N}{\lambda^2}   $ to leading order in $\lambda$. Finally, we turn to estimate the entanglement entropy from the Schmidt decomposition \eqref{eq: Schmidt decomp single conserved quantity}. To leading order
$    S_{EE}  \propto  \frac{\ln \sfrac{R}{N}}{\sfrac{R}{N} }$ \cite{suppmat}.
If $N=1$, the proportionality constant is determined:  $S_{EE}(R) = \sfrac{\ln R}{R} $ \cite{suppmat}. The analysis of the interacting fermions model \eqref{eq:FH fermions} with $N=1$ corroborates this result.  

The R\'enyi entropy is similarly inferred from the Schmidt decomposition \eqref{eq: Schmidt decomp single conserved quantity} \cite{suppmat}
\begin{equation}
        S_q(R)  \propto 
        \Bigger[10]\{\begin{array}{@{}cl}
                 \frac{1}{1-q}  (R/N)^{-q} & \text{if }0<q<1 \\[3mm]
                 \frac{q}{q-1}  \frac{1}{R/N} & \text{if }q>1.
        \end{array}
\end{equation}
For $q>1$, we find exactly $S_q(R) = \frac{q}{q-1} \frac{N}{R}$ \cite{suppmat}, corroborated for the interacting fermions model \eqref{eq:FH fermions}. 


Up to this point, the analysis was limited to a single conserved quantity. In the following, we present the proof for the power-law decay for two (or more) conserved quantities. Let us consider two conserved quantities, which is sufficient to demonstrate the difference between a single and multiple conserved quantities. 

Let the two conserved quantities be associated with the operators $\hat{N}^{i} = \hat{N}^{i} _A + \hat{N}^{i} _B $ with $i=\lbrace 1,2 \rbrace$ and where $\hat{N}^{i} _X$ acts on subsystem $X=\lbrace A,B \rbrace$.  Furthermore, the operators $\hat{N}^{1} _X,\hat{N}^{2} _X$ commute and are assumed to be lower bounded, where the bound is again set to $0$.  

Once again, assume that the perturbative analysis exists for states with fixed charges  $N_1,N_2$. The bias is defined with respect to the first charge, i.e. $R = \frac{\langle  \hat{N}^1 _A \rangle }{\langle \hat{N}^1 _B \rangle } $.

As before, the subsystem $B$ is assumed to be vacant with respect to the first charge at $\lambda \rightarrow 0$. A main difference from the single conserved quantity case arises already in the Schmidt representation of the ground state at $\lambda \rightarrow 0$:
\begin{equation}
\label{eq:G0 multiple charges}
\ket{G^{(0)}} = \sum_{k,n_2} \alpha _{n_2,k} \ket{\Phi^k _{N_1,N_2-n_2} }_A \ket{\Psi^k _{0,n_2}}_B  ,
\end{equation}
where $\ket{\Phi^k _{N_1,N-n_2} }_A$ 
and $\ket{\Psi^k _{0,n_2}}_B$ are orthonormal eigenstate sets of  $\hat{N}^{i=1,2} _{X=A,B}$ correspondingly. The subscripts of the eigenstates represent the corresponding eigenvalues.

There may be more than one Schmidt coefficient in \eqref{eq:G0 multiple charges}, implying that $S_{EE},S_q$ do not typically vanish as $\lambda\rightarrow 0$. Namely, unlike the single conserved quantity case, there may be many states with zero occupancy of $\hat{N}^1$ in the subsystem $B$. It is nevertheless interesting to study the entanglement decay, which takes a universal power-law form. Recall $\Delta S_{EE}(R) \equiv S_{EE}(R)-S_{EE}(\infty)$ and similarly define $\Delta S_q(R)$ for the R\'enyi entropy.

 $\ket{G^{(1)}} $
in its Schmidt decomposition is \cite{suppmat}
\begin{equation}
    \ket{G^{(1)}} = \sum_{n_1,n_2} \beta_{n_1,n_2,k} \ket{\phi^k _{N_1-n_1,N2_-n2}}_A \ket{\psi^k _{n_1,n2}}_B . 
\end{equation}
Here $\ket{\phi^k _{N_1-n_1,N2_-n2}}_A$ and $ \ket{\psi^k _{n_1,n2}}_B $ are orthonormal eigenstates of $\hat{N}^i _{X}$ with the subscripts denoting the eigenvalues correspondingly. Recall also that $\braket{G^{(0)}}{G^{(1)}}=0$. This fact does not suggest that 
$\Phi,\phi$ necessarily form an orthogonal set. 

First, assume that all $\phi,\Phi $ and $\Psi,\psi$ form orthonormal sets. In that case, the Schmidt coefficients are $\alpha _{n,k}$ and $|\lambda| \beta_{n_1,n_2,k} $, leading to the same behaviour at large $R$. Namely, $\Delta S_{EE}(R) \propto \ln R / R$ and $\Delta S_q (R) \propto R^{-q},1/R $ for $0<q<1$ and $q>1$ correspondingly \cite{suppmat}.

Second, assume (without loss of generality) that there is a single pair $\phi,\Phi$ that are not orthogonal. In this case, the following universal behavior is obtained \cite{suppmat}
\begin{eqnarray}
\label{eq:mltpl quantities Del S}
   \Delta  S_{EE}(R) &\propto & 1/ \sqrt{R}  \\ \nonumber 
    \Delta S_{q}(R)  &\propto& 
        \Bigger[10]\{\begin{array}{@{}cl}
                 \frac{1}{1-q}  R^{-q} & \text{if }0<q<1 \\[3mm]
                 \frac{q}{q-1}\frac{1}{\sqrt{R}} & \text{if }q>1,
                 \end{array}.
\end{eqnarray}
The essence of the proof as well as the result \eqref{eq:mltpl quantities Del S}  remain the same when more than a single pair $\phi,\Phi$ are not orthogonal in the set. Before summarising the results, it is pedagogical to demonstrate the classification of the power-law exponents for two conserved quantities in the Fermi-Hubbard model. This will illustrate that typically, not all the sets are orthogonal.

The Fermi-Hubbard model consists of spin $\sfrac{1}{2}$ fermions on a $1D$ chain of $2L$ sites. The Hamiltonian is given by 
\begin{align}
    \hat{H}_{FH} = & -t \sum^{2L-1} _{j=1,s=\uparrow \downarrow} \hat{c}^\dagger _{j+1,s} \hat{c}_{j,s} 
    + \hat{c}^\dagger _{j,s} \hat{c}_{j+1,s} 
    \\ \nonumber 
     &+ U   \sum^{2L} _{j=1}  \hat{c}^\dagger _{j,\uparrow} \hat{c}_{j,\uparrow}  
    \hat{c}^\dagger _{j,\downarrow} \hat{c}_{j,\downarrow} 
    + \mu \sum^{2L} _{j=L+1}  \hat{c}^\dagger _{j,\uparrow} \hat{c}_{j,\uparrow}  .
\end{align}
The $\mu$ term accounts for the potential offset between the subsystems, acting on the spin up fermions only. It is useful to perform once again the Jordan-Wigner transformation. Here we transform from a $1D$ chain of $2L$ sites to a $2\times 2L$ lattice, where the $\uparrow (\downarrow)$ fermions are mapped to the upper (lower) row spins \cite{reiner2016emulating}. The transformation yields 
\begin{align}
\label{eq:HQL spin half fermions}
    \hat{H}_{QL} = & -t \sum^{2L-1} _{j=1,s=\uparrow ,\downarrow} \hat{S}^+ _{j,s} \hat{S}^- _{j+1,s}  + \hat{S}^- _{j,s} \hat{S}^+ _{j+1,s}
    \\ \nonumber 
     &+ U   \sum^{2L} _{j=1}  \hat{S}^z _{j,\uparrow} \hat{S}^z _{j,\downarrow}
    + \mu \sum^{2L} _{j=L+1} \hat{S}^z _{j,\uparrow}.
\end{align}

The analysis of \eqref{eq:HQL spin half fermions} becomes quite involved at large fixed charges $N_{\uparrow,\downarrow}$. Here it will be sufficient to analyze the minimal case with $N_{\uparrow} =N_{\downarrow}=1$ with $L=1,2$, covering the range of possibilities of the ground state entanglement.



\textbf{\textit{L=1:}} Here there are two sites in each chain. The Hamiltonian is spanned by the states $\ket{k_\uparrow, k_\downarrow}$ where $k_{\uparrow,\downarrow}=1,2$ designate the site where the up and down spin states have positive eigenvalues, similarly to the interacting fermions model, now with two rows of spins. This implies $\ket{k_\uparrow, k_\downarrow}$ is an eigenvector or $\hat{S}^z _{j,s}$ with eigenvalue $    -\frac{1}{2}+\delta _{j,k_{\uparrow }}\delta _{s,\uparrow }+\delta _{j,k_{\downarrow }}\delta _{s,\downarrow }$. A perturbative analysis of $\hat{H}_{QL}/\mu  = \hat{H}_0 + \lambda \hat{V}$ can be carried out analytically \cite{suppmat}. Unlike the spinless fermions case, the state corresponding to zero spin up fermions at system $B$ is not unique. This implies that even at $R\rightarrow \infty$, both the entanglement entropy as well as the R\'enyi entropy do not vanish. Exact calculation of   $S_{EE}(\infty),S_q(\infty)$ is possible, as well as calculation of the perturbative values at large $R$, leading to \cite{suppmat}
\begin{eqnarray}
\label{eq:orthogonal exponents}
   \Delta  S_{EE}(R) &=& \frac{\ln R}{R} \\ \nonumber 
    \Delta S_{q}(R)  &=& 
        \Bigger[10]\{\begin{array}{@{}cl}
                 \frac{1}{1-q}  R^{-q} & \text{if }0<q<1 \\[3mm]
                 \frac{q}{q-1}\frac{1}{R} & \text{if }q>1.
                 \end{array}
\end{eqnarray}
Carefully following the perturbation theory reveals that indeed all the pairs $\psi,\Psi$ and $\phi,\Phi$ are orthogonal as implied by the proof.


\textbf{\textit{L=2:}}
The model is still analytically tractable. The values of $S_{EE}(\infty)$ and  $S_{q}(\infty)$ can be obtained for particular $t,U$ values at the limit $\mu\rightarrow \infty$. Writing down the perturbation theory in full glory is tedious. For this reason, the power law exponents are tested numerically. Except for the particular value $u=0$, the expected behavior of \eqref{eq:mltpl quantities Del S} is obtained. For $u=0$, the behavior \eqref{eq:orthogonal exponents} is recovered. By now, it is possible to infer orthogonality of the sets $\Psi,\psi$ and $\Phi,\phi$ correspondingly for $u=0$. This illustrates that typically one expects the full orthogonality of the sets $\Psi,\psi$ and $\Phi,\phi$ when the fixed charges are not coupled in the Hamiltonian, here $u=0$. Furthermore, it is illustrated that orthogonality of both sets $\Psi,\psi$ and $\Phi,\phi$ is the exception rather than the rule, especially when a large number of degrees of freedom is involved.



In this work, we have considered a bipartite system with either a single or multiple conserved quantities. As the system was biased to occupy subsystem $A$ with respect to one of the conserved quantities, both the entanglement entropy as well as the R\'enyi entropy of the ground state exhibit a universal power-law structure at large $R$. 

This result is in stark contrast to the mean entanglement entropy found in \cite{Page_averageEnropy,sen1996average} or to the mid-spectrum entanglement entropy \cite{haque2022entanglement}, both applicable for random states. The universal power-law structure is the result of the bias at the ground state. The excited states do not necessarily keep the large bias and hence may lack the universal structure.

A straight-forward analysis shows that a universal power-law structure can also be obtained for other entanglement quantifiers, e.g. the Concurrence  \cite{Plenio_Entanglement_Review}. It is also expected that the universal power-law structure would be apparent in the Logarithmic negativity \cite{Shpielberg_PowerLaw} and  number entanglement \cite{ma2022symmetric}, but that seems harder to prove. Nevertheless, it is possible that the results of this work could be extended to mixed states as well.  It would also be interesting to check the generality of the power-law decay for the Operationally accessible entanglement \cite{wiseman2003entanglement,Barghathi_AccessibleEnt,Barghathi_Fermions}. Finally, more work might reveal a relation between the tails of the distribution of the charge resolved entanglement \cite{goldstein2018symmetry,bonsignori2019symmetry,xavier2018equipartition} in unbiased systems to the present formalism of biased systems.

For many-body systems, it is more practical to consider entanglement witnesses rather then the entanglement entropy \cite{guhne2009entanglement}. It would be interesting to explore whether entanglement witnesses generally inherit the universal power-law structure.

Let us reiterate the importance of the universal power-law structure. First, by tunning the bias $R$, one can differentiate between a single to multiple conserved quantities from the power-law exponent of the entanglement entropy as well as the R\'enyi entropy with $q>1$. Second, for a single conserved quantity, by tunning $R$ one can infer from $\langle \hat{N}_B \rangle$ and the entanglement entropy the total occupancy $N$. Alternatively, measuring both $\hat{N}_A$ and $\hat{N}_B$ provides a tool to estimate many-body bipartite entanglement at large bias -- typically a formidable experimental challenge.   

The analysis was carried out under the assumption of a state with fixed charges. Nevertheless, when the ground state is a superposition of states with different fixed charges, the power-law behaviour remains valid. Namely, one finds $\Delta S_{EE} \propto \ln R/R$ for a single conserved quantity and $\Delta S_{EE} \propto R^{-1/2}$ for more than one conserved quantity \footnote{Complete orthogonality of the sets $\Phi,\phi$ and $\Psi,\psi$ counters this statement. Nevertheless, varying parameters arbitrarily typically destroys complete orthogonality.  }. 

The universal power-law proof relies on the existence of a perturbation theory at small $1/R$ values. In the absence of a  consistent perturbation theory, either due to a closing gap or due to nonadiabatic states, a breakdown of the power-law structure may occur. These are beyond the scope of the present work.

Acknowledgments: Dean Carmi, Shahar Hadar, Shahaf Asban, Eric Akkermans and Ofir Alon are acknowledged for interesting suggestions regarding the generality and applicability of the results.



\let\oldaddcontentsline\addcontentsline
\renewcommand{\addcontentsline}[3]{}
\bibliography{mft-meso-2}
\let\addcontentsline\oldaddcontentsline



\pagebreak

\widetext
\begin{center}
\textbf{\large Supplementary material for \\
 Universal entanglement entropy in the ground state of biased bipartite systems}

\medskip


Ohad Shpielberg$^{1,2}$  \\
\textit{\small{$^1$ Haifa Research Center for Theoretical Physics and Astrophysics, University of Haifa, Abba Khoushy Ave 199, Haifa 3498838, Israel\\
$^2$ Department of Mathematics and Physics, University of Haifa at Oranim, Kiryat Tivon 3600600,  Israel}}
\end{center}

\tableofcontents

\numberwithin{equation}{section}
\renewcommand{\thesection}{\Alph{section}} 
\renewcommand\thesubsection{\thesection.\Roman{subsection}}

\section{The Jordan-Wigner transformation and the interacting fermions model 
\label{app:sec:Jordan-Wigner}}

In this section, we recall the Jordan-Wigner transformation and use it transform the interacting fermions of $\hat{H}_{IF}$ into the spin model of   $\hat{H}_S$, up a constant shift. 

The Jordan-Wigner transformation defines the string operator $e^{i \hat{\phi}_j}$, where $\hat{\phi}_j = \pi \sum_{i<j} \hat{n}_i $ is the phase sum over the fermion occupancies to the left of $j$. This allows to define 
\begin{eqnarray}
\label{eq:JW spins}
\hat{S}^+ _j &\equiv& \hat{c}^\dagger _j  e^{i \hat{\phi}_j }
\\ \nonumber 
\hat{S}^- _j &\equiv& \hat{c} _j  e^{-i \hat{\phi}_j }
\\ \nonumber 
\hat{S}^z _j &\equiv& \frac{1}{2}\left( \hat{S}^+ _j \hat{S}^- _j - \hat{S}^- _j \hat{S}^+ _j \right) = \hat{c}^\dagger _j \hat{c}_j - \frac{1}{2}. 
\end{eqnarray}
One can check that $S^\pm _j, \hat{S}^z _j$ correspond to the spin $1/2$ operators with $\hbar=1$. From here, we find 
\begin{eqnarray}
\label{eq:JW identities}
\hat{S}^+ _j S^- _{j+1}  + \hat{S}^- _j S^+ _{j+1}  &=& \hat{f}^\dagger _j \hat{f} _{j+1} + \hat{f}^\dagger _{j+1} \hat{f} _{j}
\\ \nonumber 
\hat{S}^z _j S^z _{j+1} &=&  \frac{1}{4}   - \frac{1}{2}\left( \hat{n}_{j+1}+ \hat{n}_{j} \right) +  \hat{n}_{j+1}  \hat{n}_{j}
\\ \nonumber 
\hat{S}^z _j &=&  \hat{n}_{j} - \frac{1}{2}. 
\end{eqnarray}
Using \eqref{eq:JW identities}, we can apply the transformation from \eqref{eq:FH fermions} to 
\begin{equation}
\hat{H}_{IF} - C = \hat{H}_{S} + U \sum^{2L} _{j=1} \hat{S}^z _j. 
\end{equation}
where $C= \frac{U}{4}(2L+1) + \frac{1}{2}\mu L $. Now, noticing that $\sum^{2L} _{j=1} \langle \hat{S}^z _j \rangle = N-L $ is conserved throughout the dynamics, we can replace the $2U \sum^{2L} _{j=1}  \hat{S}^z _j $ term with the constant $2U(N-L)$. This energy shift clearly does not change the properties of the ground state when $N$ is conserved.  We are not interested in the value of the ground state energy, but rather the ground state properties. Hence, we have obtained the form of the spin Hamiltonian in \eqref{eq:FH spin Hamiltonian}. We note that the choice of adding the term $U(\hat{n}_1+ \hat{n}_{2L}) $ in \eqref{eq:FH fermions} was to allow the form \eqref{eq:FH spin Hamiltonian} without correction terms $S^z _1, S^z _{2L}$. This choice does not affect our results qualitatively, but allows for an easier analytical treatment.   



\section{Detailed calculations for the interacting fermions model with a single particle}

The purpose of this section is to present in detail the analysis of the Hamiltonian \eqref{eq:FH spin Hamiltonian} with $N=1$, i.e. the Hamiltonian \eqref{eq:Hs 1particle}. In this case, the Hamiltonian can be represented efficiently in the $\ket{k}$ basis, up to a constant shift, by  
\begin{eqnarray}
\label{eq:single particle k basis}
\hat{H}_S &=& -t\sum^{2L-1} _{k=1} \ket{k}\bra{k+1} + \ket{k+1}\bra{k} 
\\ \nonumber
&&+ \mu \sum^{2L} _{k=L+1}  \ket{k}\bra{k} + \frac{U}{2}\sum_{k=1,2L}\ket{k}\bra{k}.  
\end{eqnarray}

In the following subsections, we present the derivation of the ground state at the large offset $\mu$, resulting in large $R$ values. After obtaining the ground state, we represent it in the Schmidt decomposition 
\begin{equation}
\label{eq:GS Schmidt}
\ket{G} = \sum_k \alpha_k \ket{k}_A \ket{vac}_B + \sum_k \beta_k \ket{vac}_A \ket{k}_B , 
\end{equation}
where we recall that $\ket{k}_A \ket{vac}_B \equiv \ket{k} $ and $\ket{vac}_A \ket{k}_B \equiv \ket{k+L} $.    
 Using the Schmidt decomposition \eqref{eq:GS Schmidt}, The entanglement entropy can be written as 
 \begin{equation}
 \label{eq:HIF SEE}
 S_{EE} = - |c|^2 \ln |c|^2 -\sum_k |\beta_k|^2 \ln |\beta_k|^2 ,    
 \end{equation}
  where $|c|^2 = \sum_k |\alpha_k|^2 $.  It is also immediate to calculate the R\'enyi entropy 
  \begin{equation}
  \label{eq:HIF Sq}
  S_q = \frac{1}{1-q} \ln\left( c^q + \sum_k |\beta_k|^{2q} \right)   .     
  \end{equation}
 We recall the definitions $\lambda = -t/\mu$ and $u=-U/4t$.

\subsection{The interacting fermions model with two sites}

 Here we consider the case of $L=1$, or the system with two sites. The Hamiltonian $\hat{H}_S$ can be restricted to a $2\cross 2$ matrix 
\begin{equation}
    \hat{H}_S  /\mu = 
    \begin{pmatrix}
0 & \lambda  \\
\lambda & 1 
\end{pmatrix} 
\end{equation}
  where $\lambda= -t/\mu < 0 $ and we have shifted the Hamiltonian by a constant  $U/2$. The matrix is presented in the basis 
\begin{equation}
\ket{1} = \begin{pmatrix}
1\\ 0 
\end{pmatrix} , \quad  \ket{2} = \begin{pmatrix}
0\\ 1 
\end{pmatrix}.  
\end{equation}

The ground state of this matrix is clearly non-degenerate and gives 
\begin{equation}
    \ket{G} = \frac{1}{\sqrt{(1-\epsilon)^2+\lambda^2}} ( (1-\epsilon)\ket{1}-\lambda \ket{2}), 
\end{equation}
where $\epsilon = \frac{1-\sqrt{1+4\lambda^2}}{2}$ is the ground state energy. To attain the entanglement entropy and R\'enyi entropy, we need to write the ground state $\ket{G}$ in the Schmidt decomposition \eqref{eq:GS Schmidt}. This leads to 
\begin{equation}
    \ket{G} = \frac{1-\epsilon}{\sqrt{(1-\epsilon)^2 + \lambda^2}} \ket{1}_A \ket{vac}_B 
    -\frac{\lambda}{\sqrt{(1-\epsilon)^2 + \lambda^2}} \ket{vac}_A \ket{1}_B.
\end{equation}
At the limit of $|\lambda|\ll 1$, we find
\begin{equation}
\label{eq: 1N low lambda G}
    \ket{G}  = (1-\frac{\lambda^2}{2})\ket{1}_A\ket{vac}_B - \lambda \ket{vac}_A \ket{1}_B + O(\lambda^3).   
\end{equation}
From \eqref{eq: 1N low lambda G}, we find $R = \frac{1}{\lambda^2} + O(\lambda^4)$. Using \eqref{eq:HIF SEE}, the entanglement entropy is thus found to be   
$S_{EE}(R) =   \frac{\ln R }{R} $ to leading order. Similarly, using \eqref{eq:HIF Sq}, one can obtain the R\'enyi entropy  
\begin{equation}
        S_q(R)  = 
        \Bigger[10]\{\begin{array}{@{}cl}
                \frac{1}{1-q}R^{-q} & \text{if }0<q<1 \\[3mm]
                \frac{q}{q-1} \frac{1}{R} & \text{if }q>1.
        \end{array}
\end{equation}
Already at this simple case, it can be noted that the power law decay is achieved, with a universal exponent, independent on the details of the system.


\subsection{The interacting fermions model with four sites}

Let us turn to the analysis of the interacting fermions model, with $L=2$, i.e. four sites. The Hamiltonian is represented by the  $4\cross 4$ matrix 
\begin{equation}
\label{eq: L2 matrix}
    \hat{H}_S / \mu  = 
    \begin{pmatrix}
0 & 0 & 0 & 0  \\
0 & 0 &0 & 0   \\
0  & 0 & 1 & 0    \\
    0& 0  & 0 & 1     \\
\end{pmatrix}
    + \lambda 
    \begin{pmatrix}
2u  & 1 & 0 & 0  \\
1 & 0 & 1 & 0   \\
0  & 1 & 0 & 1    \\
0& 0  & 1 & 2u     \\
\end{pmatrix}.
\end{equation}
The matrix is in the basis 
\begin{equation}
\ket{1} = \begin{pmatrix}
1\\ 0 \\0\\0 
\end{pmatrix} ,   
\ket{2} = \begin{pmatrix}
0\\ 1 \\0\\0 
\end{pmatrix} ,  
\ket{3} = \begin{pmatrix}
0\\ 0 \\1\\0 
\end{pmatrix} ,
\ket{4} = \begin{pmatrix}
0\\ 0 \\0\\1 
\end{pmatrix} 
.  
\end{equation}
Since we aim to analyze the large $\mu$ behaviour (small $|\lambda|$), we consider the perturbative approach: 
\begin{equation}
\frac{1}{\mu} \hat{H}_S = \hat{H}_0 + \lambda \hat{V}  ,    
\end{equation}
 where $\hat{H}_0, \hat{V}$ refer  to the first and second matrices in  \eqref{eq: L2 matrix}. 
Clearly, at $\lambda=0$, the ground state is degenerate. The standard degenerate perturbation theory leads to the ground state 
$\ket{G} = \beta_1 \ket{3} + \sum^2 _{k=1} \alpha_k \ket{k}   + O(\lambda^2)$ where 
\begin{eqnarray}
\alpha_1 &=& \frac{1}{\mathcal{N}}\left( \frac{\epsilon}{\sqrt{1+\epsilon^2}} + \lambda \frac{2u-\epsilon}{2(u-\epsilon)}\frac{1}{1+(2u-\epsilon)^2} \frac{1}{\sqrt{1+\epsilon^2}}
\right)
 \nonumber \\ 
 \alpha_2 & = &
 \frac{1}{\mathcal{N}}\left(
 \frac{1}{\sqrt{1+\epsilon^2}}+\lambda \frac{1}{\sqrt{1+\epsilon^2}} \frac{1}{2(u-\epsilon)(1+(2u-\epsilon)^2)}
 \right)
 \\ \nonumber 
 \beta_1 & = & -\frac{\lambda}{\sqrt{1+\epsilon^2}}
 \frac{1}{ \mathcal{N} } ,  
 \end{eqnarray}
and $1 =   \sqrt{\alpha^2 _1 + \alpha^2 _2 + \beta^2 _1} $ is a normalization factor. Here, $\epsilon = u -\sqrt{1+u^2}$ is the ground state energy at $\lambda\rightarrow 0$. We can represent the ground state in the Schmidt decomposition 
\begin{equation}
    \ket{G} = \sqrt{\alpha^2 _1 + \alpha^2 _2} \left( \frac{\alpha_1}{\sqrt{\alpha^2 _1 + \alpha^2 _2}} \ket{1}_A  + \frac{\alpha_2}{\sqrt{\alpha^2 _1 + \alpha^2 _2}} \ket{2}_A  \right) \ket{vac}_B + \beta_1 \ket{vac}_A \ket{1}_B. 
\end{equation}
To leading order, we find that $R = \frac{1+\epsilon^2}{\lambda^2}$. At the limit of small $|\lambda|$ (large $R$), we find   $ S_{EE}(R) = \frac{\ln R}{R}$ to leading order from \eqref{eq:HIF SEE}. Similarly, one can obtain the R\'enyi entropy from \eqref{eq:HIF Sq} 
\begin{equation}
        S_q(R)  = 
        \Bigger[10]\{\begin{array}{@{}cl}
                \frac{1}{1-q}R^{-q} & \text{if }0<q<1 \\[3mm]
                \frac{q}{q-1} \frac{1}{R} & \text{if }q>1.
        \end{array}
\end{equation}
Namely, even though we have changed the system size from $L=1$ to $L=2$, both the entanglement entropy and the R\'enyi entropy are independent of $u$ at the large $R$ limit. Moreover, we notice that the power law behaviour is universal as the exponent is independent of the parameters of the model at the large $R$ limit.


\subsection{Numerical treatment of the interacting fermions model with a single fermion}

For $L>2$, the analytical perturbative approach becomes cumbersome. Moreover, for $N=1$, it was ascertained in the main text that the entanglement entropy as well as the R\'enyi entropy attain a universal structure, independent of the model parameters. 

For this reason, analysis of the $L>2$ cases is performed numerically. Namely, the ground state is found by exact diagonalization of the matrix Hamiltonian  \eqref{eq:single particle k basis} -- a $2L\cross 2L$ matrix. After having obtained the ground state at a particular parameter set $(U,t,L)$ and for a range of $\mu$ values selected to obtain the large $R$ limit, it is possible to extract numerically the entanglement entropy and the R\'enyi entropy. The results are presented in Fig.~\ref{fig:IF N=1 universal plot}.

\section{The interacting fermions model with two fermion particles}

To directly demonstrate that the results hold also for $N>1$, we consider the interacting fermions model with $N=2$. As before, we write the Hamiltonian in its reduced matrix form and find its ground state. Then, computing the values of $R(\lambda)$ as well as $S_{EE},S_q$, we show they take the announced power-law form. 

First, we denote by $\ket{k_1,k_2}$ the position of the particles, where $1\leq k_1 < k_2 \leq 2L$. Note that  $S^z _j \ket{k_1,k_2} =\left( -\frac{1}{2}+\delta_{j,k_1}+\delta_{j,k_2} \right)\ket{k_1,k_2} $. In what follows, we span vectors in the basis 
\begin{equation}
{\ket{1,2},\ket{1,3},...,\ket{1,2L},\ket{2,3},...,\ket{2,2L},...,\ket{2L-1,2L}}. 
\end{equation}
Notice that the dimension of the vector is $L (2L-1) $

\subsection{The interacting fermions model with two fermions on four sites }

Let us consider now the interacting fermions model with $N=2,L=2$. In this case, we can represent the Hamiltonian by the $6\times6$ matrix $\frac{1}{\mu} \hat{H}_S = \hat{H}_0 + \lambda \hat{V} $ where 
\begin{equation}
\hat{H_0} = 
    \begin{pmatrix}
    0 & 0 & 0  & 0  & 0 & 0 \\
    0 & 1 & 0  & 0  & 0 & 0 \\
    0 & 0 & 1  & 0  & 0 & 0 \\
    0 & 0 & 0  & 1  & 0  & 0 \\
    0 & 0 & 0  & 0  & 1 & 0 \\
    0 & 0 & 0  & 0  & 0 & 2  
    \end{pmatrix}, 
    \hat{V} = 
    \begin{pmatrix}
    u & 1   & 0  & 0  & 0   & 0 \\
    1 & -3u & 1  & 1  & 0   & 0 \\
    0 & 1   & -u & 0  & 1   & 0 \\
    0 & 1   & 0  & -u & 1   & 0 \\
    0 & 0   & 1  & 1  & -3u & 1 \\
    0 & 0   & 0  & 0  & 1   & u  
    \end{pmatrix}. 
\end{equation}
The matrix is in the basis 
\begin{equation}
\ket{1,2} = \begin{pmatrix}
1\\ 0 \\0\\0 \\0\\0 
\end{pmatrix} ,   
\ket{1,3} = \begin{pmatrix}
0\\ 1 \\0\\0 \\0\\0 
\end{pmatrix} ,  
\ket{1,4} = \begin{pmatrix}
0\\ 0 \\1\\0 \\0\\0 
\end{pmatrix} ,
\ket{2,3} = \begin{pmatrix}
0\\ 0 \\0\\1 \\0\\0 
\end{pmatrix} 
\ket{2,4} = \begin{pmatrix}
0\\ 0 \\0\\0 \\1\\0 
\end{pmatrix} 
\ket{3,4} = \begin{pmatrix}
0\\ 0 \\0\\0 \\0\\1 
\end{pmatrix} 
.  
\end{equation}

 Clearly, at the large $\mu$ (small $|\lambda|$) limit the ground state is non-degenerate. We use the non-degenerate perturbation theory to write the ground state to order $\lambda$:
 \begin{eqnarray}
 \label{eq:gs for N=2 L=2}
     \ket{G(\lambda} &=& \frac{1}{\sqrt{1+4\lambda^2}}\left(\ket{1,2} -2 \lambda \ket{1,3} \right)  + O(\lambda^2) 
     \\ \nonumber 
     &=& \frac{1}{\sqrt{1+4\lambda^2}}\left(\ket{1,2}_A \ket{vac}_B -2 \lambda \ket{1}_A\ket{1}_B \right)  + O(\lambda^2) 
 \end{eqnarray}
From \eqref{eq:gs for N=2 L=2}, we find $1/R = 2\lambda^2 $
and the entanglement entropy $S_{EE}(R) =  \frac{\ln R/2}{R/2}$, both to leading order. For the R\'enyi entropy, we recover 
\begin{equation}
        S_q(R)  = 
        \Bigger[10]\{\begin{array}{@{}cl}
                \frac{1}{1-q}(R/2)^{-q} & \text{if }0<q<1 \\[3mm]
                \frac{q}{q-1} \frac{1}{R/2} & \text{if }q>1.
        \end{array}
\end{equation}
For the interacting fermions model, we can thus conjecture that for $N$ fermion particles, we find 
\begin{eqnarray}
\label{eq:universal entanglement N=2}
S_{EE}(R) &=& \frac{\ln R/N}{R/N}
\\ \nonumber 
        S_q(R)  &=& 
        \Bigger[10]\{\begin{array}{@{}cl}
                \frac{1}{1-q}(R/N)^{-q} & \text{if }0<q<1 \\[3mm]
                \frac{q}{q-1} \frac{1}{R/N} & \text{if }q>1.
        \end{array}
\end{eqnarray}
Here the system size does not play a role as long as $L\geq N$. In the following subsection, we numerically verify this conjecture for $L>2$ and $N=2$.



\subsection{Numerical treatment of the interacting fermions model with two fermion particles}

\begin{figure}
    \centering
    \includegraphics[scale=0.80]{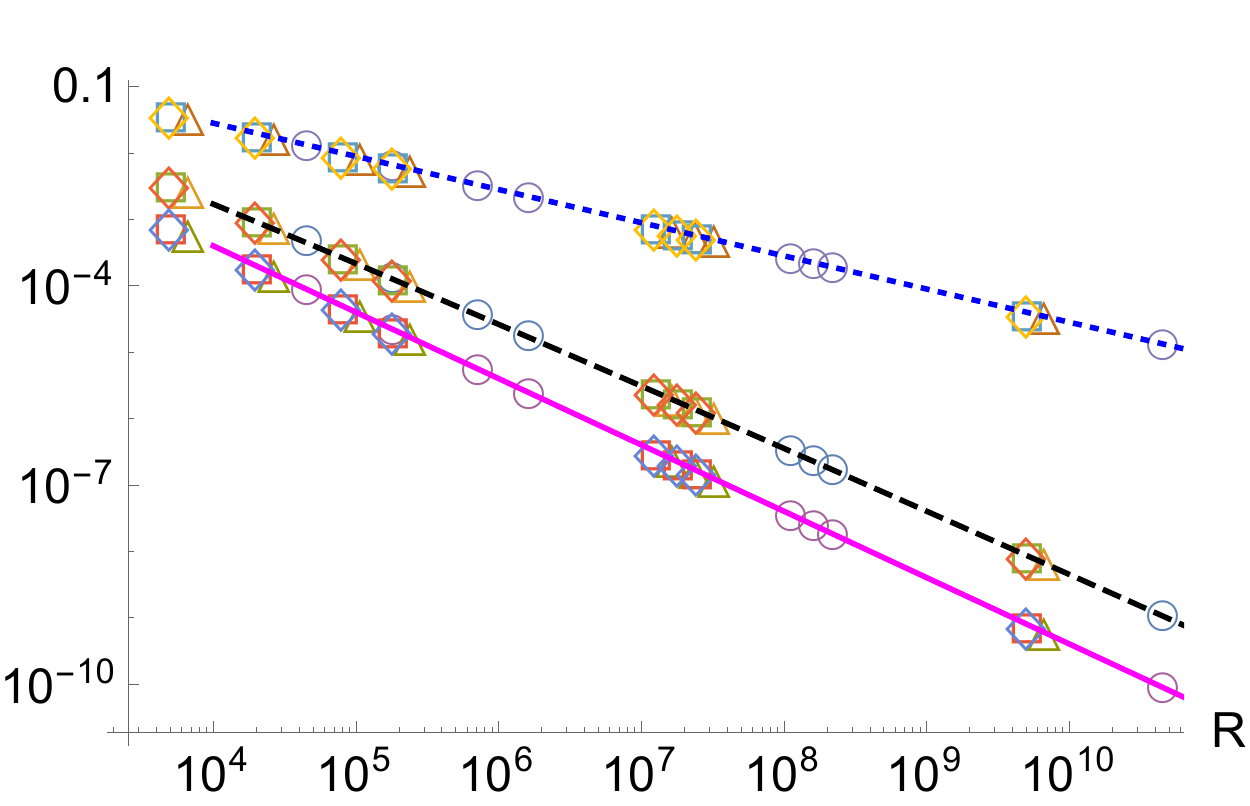}
    \caption{ The entanglement entropy and R\'enyi entropy with $q=\frac{1}{2},2$ are numerically calculated in the range   $\mu \in \left[10^2,10^5 \right]$ and for four parameter sets $(U,t,L)$ with $N=2$.
Circles: $(1,1,2)$; triangles: $(0,1,3)$; Squares: $(\frac{1}{2},1,4)$; Diamonds: $(2,1,4)$. All the points attain large $R$ values and collapse onto the universal curves $S_{EE} = \frac{\ln (R/2)}{R/2}$ (black dashed line), $S_q = \frac{1}{1-q}(R/2)^{-q}$ (magenta solid line) for $0<q=\frac{1}{2}<1$ and $S_q = \frac{q}{q-1}\frac{1}{R/2}$ (blue dotted line) for $1<q=2$. Note that no fitting parameters are required, further validating the analytical predictions.   }
    \label{fig:IF N=2 universal plot}
\end{figure}

For $L>2$, the analytical perturbative approach becomes cumbersome. Moreover, a universal form was conjectured in \eqref{eq:universal entanglement N=2}. The analysis of the $L>2$ cases is thus performed numerically. Namely, the ground state is found by exact diagonalization of the matrix Hamiltonian $\hat{H}_S$ with $N=2$ particles -- a $L(2L-1)\cross L(2L-1)$ matrix. After having obtained the ground state at a particular parameter set $(U,t,L)$ and for a range of $\mu$ values selected to obtain the large $R$ limit, it is possible to extract numerically the entanglement entropy and the R\'enyi entropy. The results are presented in Fig.~\ref{fig:IF N=2 universal plot}.

\section{Entanglement of a single conserved quantity: auxiliary proofs \label{app:sec N1 aux proofs}}

In the main text, the universal power-law at large $R$ relies on the Schmidt decomposition of $\ket{G^{(1)}}$. Here, it is shown that the Schmidt decomposition of $\ket{G^{(1)}}$ can indeed be written using the orthonormal sets of $\hat{N}_A$ and $\hat{N}_B$ eigenstates. Moreover, we bring in full detail the calculations of the entanglement entropy and the R\'enyi entropy based on the Schmidt decomposition of the ground state to first order in $\lambda$.

\subsection{The Schmidt decomposition of the first order state }

The state $\ket{G^{(1)}}$ lies in the subspace of the Hilbert space, with fixed charge $N$ in the complement to the projector defined by $\ket{G^{(0)}}\bra{G^{(0)}}$. Therefore, there exist  $\ket{\tilde{\phi}^{k'} _{n_A}}_A,\ket{\psi^{k} _{n_B}}_B$ orthonormal sets of eigenstates of $\hat{N}_A$ and $\hat{N}_B$ with the corresponding eigenvalues  $n_A$ and $n_B$. Namely, we can write 
\begin{equation}
\label{app:eq:G1 orthonormal decomp}
    \ket{G^{(1)}} = \sum_{k,k'}\sum^N _{n\geq 0}  \gamma_{k,k',n} \ket{\tilde{\phi}^{k'} _{N-n}}_A \ket{\psi^{k} _{n}}_B.
\end{equation}
Without loss of generality, we assume \eqref{app:eq:G1 orthonormal decomp} is normalized. 
Then, we can rewrite \eqref{app:eq:G1 orthonormal decomp} as 
\begin{equation}
\label{app:eq:G1 Schmidt decomp}
    \ket{G^{(1)}} = \sum^N _{n\geq 0} \sum_k \alpha_{n,k}
\ket{\phi^{k} _{N-n}}_A \ket{\psi^{k} _{n}}_B, 
\end{equation}
where $\alpha_{n,k} \ket{\phi^k _{N-n} }_A = \sum_{k'} \gamma_{k,k',n} \ket{\tilde{\phi}^k _{N-n} }_A$ and $|\alpha_{n,k}|^2  = \sum_{k'} |\gamma_{k,k',n}|^2  $. It is easy to verify that  $\ket{\phi^{k} _{N-n}}_A$ are orthonormal themselves. Up to an irrelevant absorption of phase in the states, one can always recover $\beta_{k,n}$ as non-negative reals. Therefore \eqref{app:eq:G1 Schmidt decomp} is indeed written in the Schmidt decomposition.

\subsection{Detailed calculation of the entanglement entropy and the R\'enyi entropy }

Let us consider the Schmidt decomposition of the ground state $\ket{G(\lambda)}$ to first order in $\lambda$
\begin{eqnarray}
\label{app:eq: Schmidt decomp single conserved quantity}
     \ket{G(\lambda)} &=& \frac{1}{\sqrt{\mathcal{N}_\lambda} } \ket{\Phi^\lambda _N} _A \ket{vac}_B 
       \\ \nonumber 
     &&
     +\frac{\lambda}{\sqrt{\mathcal{N}_\lambda} } \sum^N _{n> 0,k} \alpha_{n,k} \ket{\phi^{k} _{N-n}}_A \ket{\psi^{k} _{n}}_B,  \nonumber \\  \nonumber
     \ket{\Phi^\lambda _N}_A &=& \ket{\Phi_N}_A + \lambda \sum_k \alpha_{0,k}  \ket{\phi^k _N}.
\end{eqnarray}
First, let us use \eqref{app:eq: Schmidt decomp single conserved quantity} to calculate $R$ to leading order: 
\begin{eqnarray}
    R = \frac{ \bra{G(\lambda)} \hat{N}_A \ket{G(\lambda)} }{ \bra{G(\lambda)} \hat{N}_B \ket{G(\lambda)}} = \frac{ N + N \lambda^2 \sum_k |\alpha_{0,k}|^2 + \lambda^2 \sum_{k,n>0} (N-n) |\alpha_{n,k}|^2 + O(\lambda^3) }{ \lambda^2 \sum_{k,n>0} n |\alpha_{n,k}|^2 +  O(\lambda^3)}
    \propto \lambda^{-2}.
\end{eqnarray}
In particular when $N=1$, we find to leading order $R = \frac{1}{\lambda^2 b}$, where $b= \sum_{k,n>0}   |\alpha_{n,k}|^2$. 

To calculate the entanglement entropy, we use the Schmidt coefficients in eq.\eqref{app:eq: Schmidt decomp single conserved quantity}
\begin{equation}
    S_{EE}(\lambda) = - \frac{1+\lambda^2 (c-b)  }{1+c\lambda^2} \ln \frac{1+\lambda^2 (c-b) }{1+c\lambda^2}  
      - \frac{\lambda^2}{1+c\lambda^2} \sum_{k,n>0} |\alpha_{n,k}|^2 \ln \frac{\lambda^2}{1+c\lambda^2} |\alpha_{n,k}|^2 + O(\lambda^3),  
\end{equation}
where $c = \sum_{k,n\geq 0} |\alpha_{n,k}|^2$. A simple perturbative analysis leads to 
\begin{equation}
    S_{EE}(\lambda) =  \lambda^2 ( b - \sum_{k,n>0} |\alpha_{k,n}|^2 \ln |\alpha_{k,n}|^2 - b \ln \lambda^2  ) = - b \lambda^2 \ln \lambda^2 + O(\lambda^2) . 
\end{equation}
This implies that to leading order $S_{EE}(R) \propto \frac{\ln R}{R}   $. In particular, for $N=1$, we find $S_{EE} = \frac{\ln R}{R} $ to leading order. Namely, we have recovered the proportionality coefficient.   

Lastly, we find the R\'enyi entropy from the Schmidt coefficients in eq.\eqref{app:eq: Schmidt decomp single conserved quantity}
\begin{equation}
    S_q(\lambda) = \frac{1}{1-q} \ln \left( \frac{(1+\lambda^2 (c-b))^q }{(1+c\lambda^2)^q}  + \frac{\lambda^{2q}}{(1+c\lambda^2)^q}  \sum_{k,n>0} |\alpha_{n,k}|^{2q}
    \right) + O(\lambda^3) .
\end{equation}
To leading order, we find 
\begin{equation}
    S_q(\lambda) = \frac{1}{1-q} \ln \left( 1- b q \lambda^2  + 
      \lambda^{2q} \sum_{k,n>0} |\alpha_{n,k}|^{2q}
    \right) + O(\lambda^3) .
\end{equation}
We can recover the leading order behaviour for either $0<q<1$ or $q>1$
\begin{equation}
        S_q(R)  \propto 
        \Bigger[10]\{\begin{array}{@{}cl}
                 \frac{1}{1-q}  R^{-q} & \text{if }0<q<1 \\[3mm]
                 \frac{q}{q-1}\frac{1}{R} & \text{if }q>1,
        \end{array}
\end{equation}
In particular, for $q>1$, we find $S_q(R)= \frac{q}{q-1} \frac{N}{R}$. We notice that for $0<q<1$ the prefactor of the leading order $S_q(R)$ is not generally determined. Only for the particular case, where to first order the is a single non-zero $\alpha_{n,k}$ (as it happens for the interacting fermions model), the prefactor is determined such that $S_q(R) = \frac{1}{1-q}\left(\frac{N}{R}\right)^q $.


\section{Two conserved fields: The Fermi-Hubbard model with spin half fermions  }

The purpose of this section is to present in detail the entanglement entropy and R\'enyi entropy analysis of the Fermi-Hubbard model on a $1D$ chain of $2L$ sites, occupied by spin $\sfrac{1}{2}$ fermions. Assume the simplest non-trivial case of $N_{\uparrow} = N_{\downarrow} =1 $ where there is a single spin up and spin down fermions. In this setup, the Fermi-Hubbard Hamiltonian is given by  
\begin{eqnarray}
    \hat{H}_{QL} &=&  \sum^{2L} _{k_\uparrow, k_\downarrow = 1} \left(  U \delta_{k_\downarrow,k_\uparrow} + \mu \, \Theta \left( k_\uparrow -L - \frac{1}{2} \right) \right) \ket{k_\uparrow, k_\downarrow}\bra{k_\uparrow, k_\downarrow}
    \\ \nonumber 
    && -t \sum^{2L-1} _{k_\uparrow = 1 } \sum^{2L} _{k_\downarrow=1}
    \ket{k_\uparrow+1, k_\downarrow}\bra{k_\uparrow, k_\downarrow} +\ket{k_\uparrow, k_\downarrow}\bra{k_\uparrow+1, k_\downarrow}
    \\ \nonumber 
    && -t \sum^{2L-1} _{k_\downarrow = 1 } \sum^{2L} _{k_\uparrow=1}
    \ket{k_\uparrow, k_\downarrow+1}\bra{k_\uparrow, k_\downarrow} +\ket{k_\uparrow, k_\downarrow}\bra{k_\uparrow+1, k_\downarrow+1}.
\end{eqnarray}
Here $\Theta (x)$ is the Heaviside function. In the next subsections, we present the derivation of the ground state at large $\mu$ bias, leading to the large $R$ limit. After obtaining the ground state, we represent it in the Schmidt decomposition 
\begin{eqnarray}
    \ket{G} &=& \sum_{\substack{k_\uparrow <L+1  \\ k_\downarrow<L+1}}  \alpha_{k_\uparrow,k_\downarrow} \ket{k_\uparrow,k_\downarrow}_A \ket{vac}_B + 
     \sum_{\substack{k_\uparrow <L+1  \\ k_\downarrow \geq L+1}}\alpha_{k_\uparrow,k_\downarrow} \ket{k_\uparrow}_A \ket{k_\downarrow - L}_B 
    \\ \nonumber 
     && +  
    \sum_{\substack{k_\uparrow \geq L+1  \\ k_\downarrow<L+1}}\alpha_{k_\uparrow,k_\downarrow} \ket{k_\downarrow}_A \ket{k_\uparrow-L}_B 
    +
    \sum_{\substack{k_\uparrow  \geq L+1  \\ k_\downarrow \geq L+1}}\alpha_{k_\uparrow,k_\downarrow} \ket{vac}_A \ket{k_\downarrow-L,k_\uparrow-L}_B ,
\end{eqnarray}
where $\ket{k_s}_{X}$ indicates there is a single up pointing spin in row $s$ and position $k$ in the subsystem $X=\lbrace A,B \rbrace$. Moreover, we assume the state is normalized, so $\sum_{k_\uparrow,k_\downarrow} | \alpha_{k_\uparrow,k_\downarrow} |^2 = 1$.  The Schmidt decomposition implies 
\begin{equation}
    S_{EE} = - (\sum_{k_\uparrow,k\downarrow \in \mathcal{B}}  |\alpha_{k_\uparrow,k_\downarrow}|^2) \ln (\sum_{k_\uparrow,k\downarrow \in \mathcal{B}}  |\alpha_{k_\uparrow,k_\downarrow}|^2) - \sum_{k_\uparrow,k_\downarrow \notin \mathcal{B}} |\alpha_{k_\uparrow,k_\downarrow}|^2 \ln |\alpha_{k_\uparrow,k_\downarrow}|^2),
\end{equation}
where $k_{\uparrow,\downarrow} \in \mathcal{B}$ if $k_s < L+1$ for both $s=\uparrow,\downarrow$. Similarly, the R\'enyi entropy is given by 
\begin{equation}
    S_q =\frac{1}{1-q}
    \ln \left(  (\sum_{k_\uparrow,k_\downarrow \in \mathcal{B}}  |\alpha_{k_\uparrow,k_\downarrow}|^2)^q 
    +  \sum_{k_\uparrow,k_\downarrow \notin \mathcal{B}} |\alpha_{k_\uparrow,k_\downarrow}|^{2q} 
    \right). 
\end{equation}
We remind that $\lambda = -t/\mu$ and $u=-U/4t$, as has been defined throughout the work. 


\subsection{The Fermi-Hubbard model with two sites}

  Here we present in detail the analysis of the ground state of $L=1$ with $N_\uparrow = N_\downarrow=1$ of the Fermi-Hubbard model at large bias. In this case, we have four relevant states $\ket{k_\uparrow,k_\downarrow} = \lbrace \ket{1,1},\ket{1,2},\ket{2,1},\ket{2,2}  \rbrace $. In this basis, the Hamiltonian is restricted to the  $4\cross 4 $ matrix 
\begin{equation}
    \hat{H}_{QL}  /\mu =  \hat{H}_0 + \lambda \hat{V} = 
    \begin{pmatrix}
0 & 0 & 0 & 0   \\
0 & 0 & 0 & 0   \\
0 & 0 & 1 & 0   \\
0&0&0 & 1 
\end{pmatrix} 
 + \lambda 
 \begin{pmatrix}
4u & 1 & 1 & 0   \\
1 & 0 & 0 & 1   \\
1 & 0 & 0 & 1   \\
0&1&1 & 4u 
\end{pmatrix} .
\end{equation}
   At $\lambda =  0$, the ground state is doubly degenerate. The standard first order perturbation theory leads to  
  \begin{eqnarray}
      \ket{G} &=&  \alpha_{1,1} \ket{1,1}_A \ket{vac}_B
       +\alpha_{1,2} \ket{1_\uparrow}_A \ket{1_\downarrow}_B+ \alpha_{2,1} \ket{1_\downarrow}_A \ket{1_\uparrow}_B +  \alpha_{2,2} \ket{vac}_A \ket{1,1}_B
       + O(\lambda^3) , 
      \\ \nonumber 
      && 
      \begin{matrix}
       \alpha_{1,1} = \frac{\epsilon -\frac{\lambda  (4 u-\epsilon ) \left(4 u \epsilon -\epsilon ^2+1\right)}{(2 \epsilon -4 u) \left((\epsilon -4 u)^2+1\right)}}{\sqrt{\lambda ^2+1} \sqrt{\epsilon ^2+1}} 
       & \alpha_{2,1} = -\frac{\lambda  \epsilon }{\sqrt{\lambda ^2+1} \sqrt{\epsilon ^2+1}}  \\
       \alpha_{1,2}  =  \frac{1-\frac{\lambda  \left(4 u \epsilon -\epsilon ^2+1\right)}{(2 \epsilon -4 u) \left((\epsilon -4 u)^2+1\right)}}{\sqrt{\lambda ^2+1} \sqrt{\epsilon ^2+1}} &
       \alpha_{2,2} =  -\frac{\lambda }{\sqrt{\lambda ^2+1} \sqrt{\epsilon ^2+1}},
      \end{matrix}
  \end{eqnarray}
where the ground state energy is $\epsilon = 2u - \sqrt{1+4u^2}$.  To leading order we find $R= 1/\lambda^2$ and 
\begin{eqnarray}
    S_{EE}(R)-S_{EE}(\infty) &=& \frac{\ln R}{R} 
    \\ \nonumber 
  S_q(R)  =  
        \Bigger[10]\{\begin{array}{@{}cl}
                 \frac{1}{1-q}  R^{-q} & \text{if }0<q<1 \\[3mm]
                 \frac{q}{q-1}\frac{1}{R} & \text{if }q>1,
        \end{array}
\end{eqnarray}
where 
\begin{eqnarray}
S_{EE}(\infty) &=& -\frac{\epsilon ^2 \ln \left(\frac{\epsilon }{\sqrt{\epsilon ^2+1}}\right)-\ln \left(\epsilon ^2+1\right)}{\epsilon ^2+1}.
\\ \nonumber 
S_q (\infty) &=&  \frac{1}{1-q} \ln \left( 
(\frac{\epsilon^2}{1+\epsilon^2})^q + (\frac{1}{1+\epsilon^2})^q
\right).
\end{eqnarray}

The above $L=1$ results agree with the single conserved quantity scenario. However, as explained in the main text, the entanglement typically attains a different power-law exponents. To observe the typical case, we explore the Fermi-Hubbard model with four sites ($L=2$) in the next section.


\subsection{The Fermi-Hubbard model with four sites}
\begin{figure}
    \centering
    \includegraphics[scale=0.420]{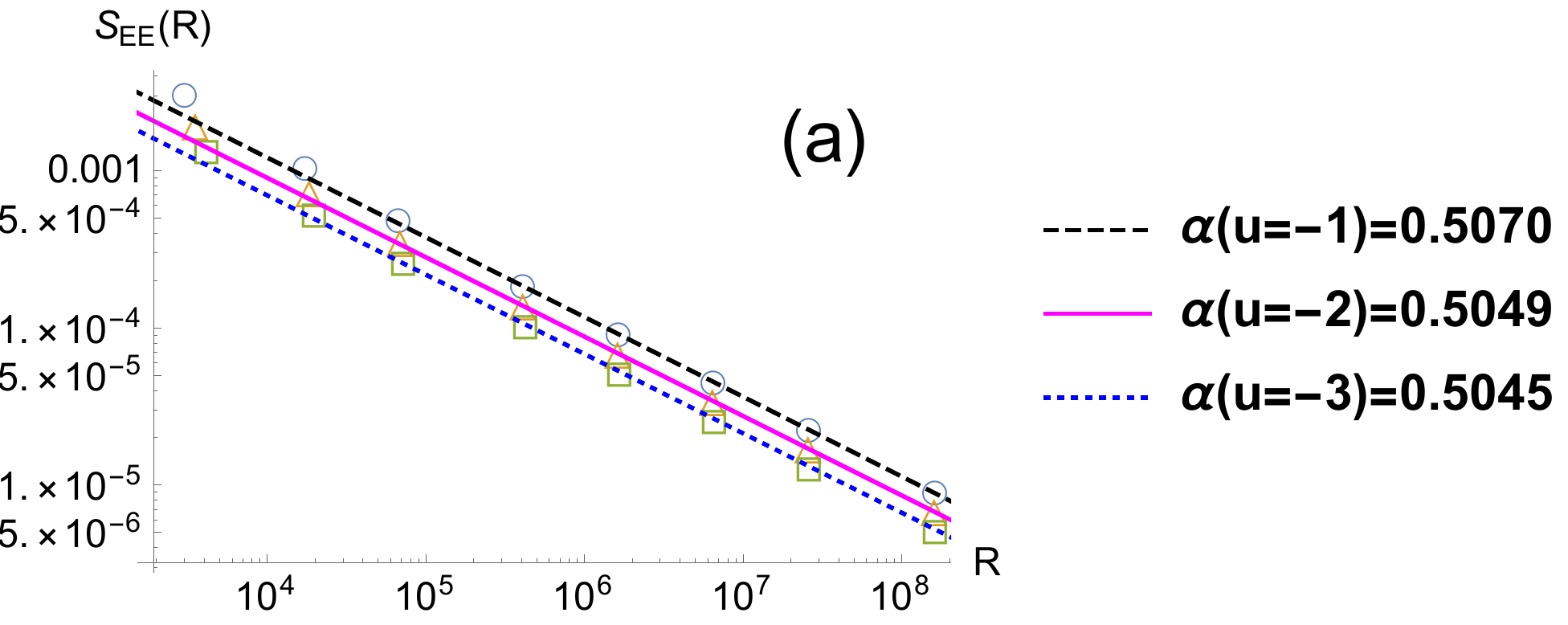}
    \includegraphics[scale=0.420]{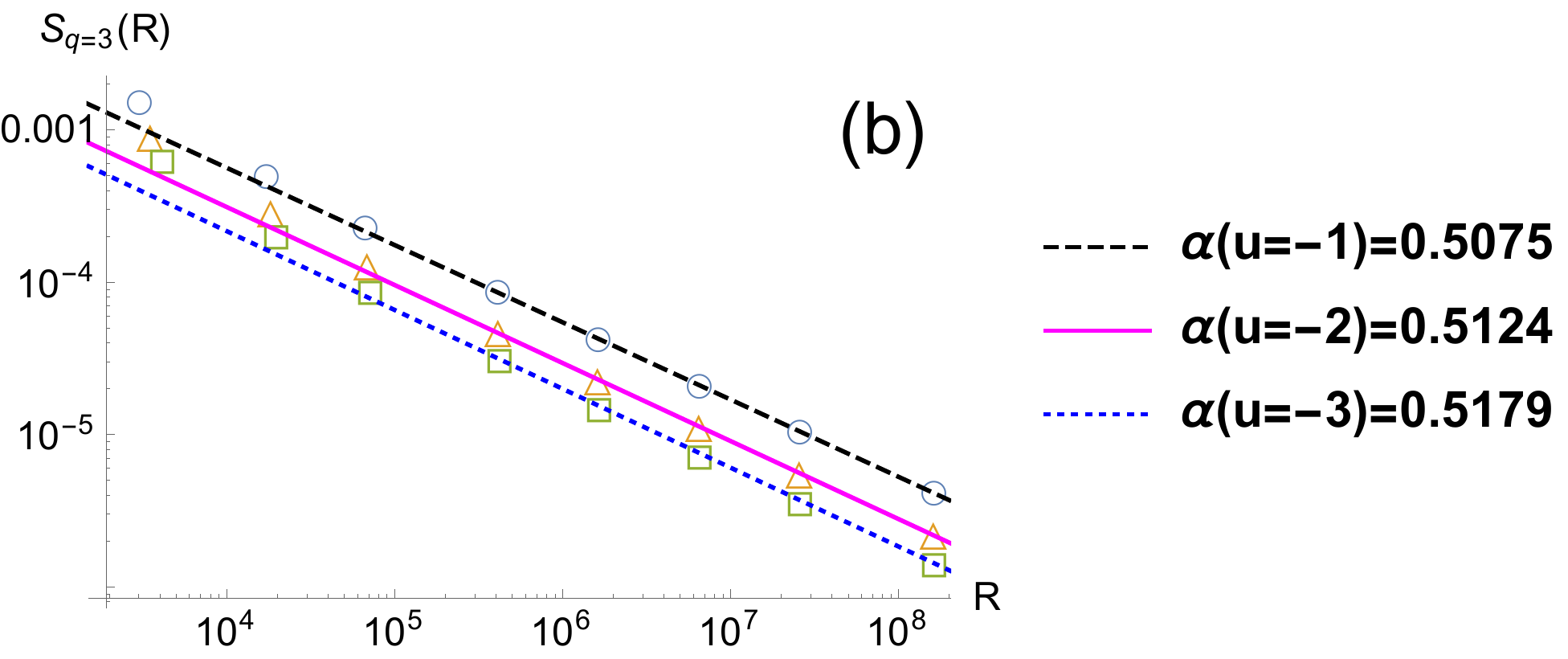}
    \includegraphics[scale=0.42]{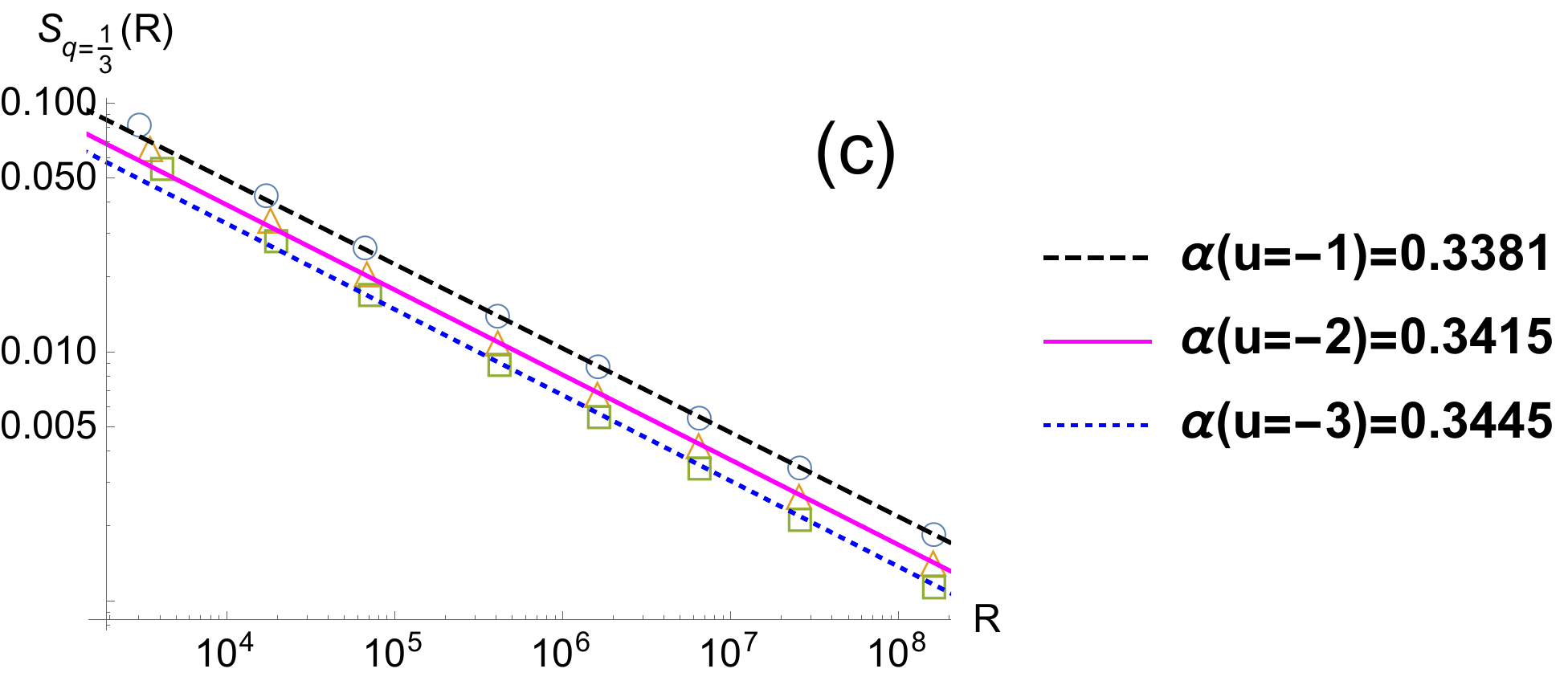}
    \caption{ For the Fermi-Hubbard model, the entanglement entropy (a) and the R\'enyi entropy with $q=3,\frac{1}{3}$ (b) and (c) correspondingly are numerically calculated in the range   $\lambda \in \left[10^{-4},10^{-1} \right]$ with $t=1,L=2$ and for the $u$ parameters $u=\lbrace -1,-2,-3\rbrace$ (circles, triangles and squares correspondingly). 
  All the points attain large $R$ values and are fitted with good agreement to the analytical predictions;   $\Delta S_{EE}(R) \propto  \sfrac{1}{\sqrt{R}}$,
  $\Delta S_{q>1}(R) \propto \sfrac{1}{\sqrt{R}}$ and  
  $\Delta S_{0<q<1}(R) \propto \sfrac{1}{R^q}$. The legend $\alpha$ indicate the fitted exponent values for the different plots. 
Note that the axes are plotted in logarithmic scales to highlight the power-law structure of the entanglement.  }
    \label{fig:F-H  Nup=Ndown=1 plot}
\end{figure}

  Here we present in detail the analysis of the ground state of $L=2$ with $N_\uparrow = N_\downarrow=1$ of the Fermi-Hubbard model at large bias. In this case, we have sixteen  relevant states 
  \begin{equation}
  \ket{k_\uparrow,k_\downarrow} = \lbrace \ket{1,1},\ket{1,2},\ket{1,3},\ket{1,4}
  ,\ket{2,1},\ket{2,2},\ket{2,3},\ket{2,4}
  ,\ket{3,1},\ket{3,2},\ket{3,3},\ket{3,4}
  ,\ket{4,1},\ket{4,2},\ket{4,3},\ket{4,4}
   \rbrace .    
  \end{equation}
   In this basis, the Hamiltonian is restricted to the  $16\cross 16 $ matrix, where 
   $\hat{H}_{QL}  /\mu =  \hat{H}_0 + \lambda \hat{V}  $ and 
\begin{equation}
\label{app:eq:16 matrix}
    \hat{H}_0 = \left(
\begin{array}{cccccccccccccccc}
 0 & 0 & 0 & 0 & 0 & 0 & 0 & 0 & 0 & 0 & 0 & 0 & 0 & 0 & 0 & 0 \\
 0 & 0 & 0 & 0 & 0 & 0 & 0 & 0 & 0 & 0 & 0 & 0 & 0 & 0 & 0 & 0 \\
 0 & 0 & 0 & 0 & 0 & 0 & 0 & 0 & 0 & 0 & 0 & 0 & 0 & 0 & 0 & 0 \\
 0 & 0 & 0 & 0 & 0 & 0 & 0 & 0 & 0 & 0 & 0 & 0 & 0 & 0 & 0 & 0 \\
 0 & 0 & 0 & 0 & 0 & 0 & 0 & 0 & 0 & 0 & 0 & 0 & 0 & 0 & 0 & 0 \\
 0 & 0 & 0 & 0 & 0 & 0 & 0 & 0 & 0 & 0 & 0 & 0 & 0 & 0 & 0 & 0 \\
 0 & 0 & 0 & 0 & 0 & 0 & 0 & 0 & 0 & 0 & 0 & 0 & 0 & 0 & 0 & 0 \\
 0 & 0 & 0 & 0 & 0 & 0 & 0 & 0 & 0 & 0 & 0 & 0 & 0 & 0 & 0 & 0 \\
 0 & 0 & 0 & 0 & 0 & 0 & 0 & 0 & 1 & 0 & 0 & 0 & 0 & 0 & 0 & 0 \\
 0 & 0 & 0 & 0 & 0 & 0 & 0 & 0 & 0 & 1 & 0 & 0 & 0 & 0 & 0 & 0 \\
 0 & 0 & 0 & 0 & 0 & 0 & 0 & 0 & 0 & 0 & 1 & 0 & 0 & 0 & 0 & 0 \\
 0 & 0 & 0 & 0 & 0 & 0 & 0 & 0 & 0 & 0 & 0 & 1 & 0 & 0 & 0 & 0 \\
 0 & 0 & 0 & 0 & 0 & 0 & 0 & 0 & 0 & 0 & 0 & 0 & 1 & 0 & 0 & 0 \\
 0 & 0 & 0 & 0 & 0 & 0 & 0 & 0 & 0 & 0 & 0 & 0 & 0 & 1 & 0 & 0 \\
 0 & 0 & 0 & 0 & 0 & 0 & 0 & 0 & 0 & 0 & 0 & 0 & 0 & 0 & 1 & 0 \\
 0 & 0 & 0 & 0 & 0 & 0 & 0 & 0 & 0 & 0 & 0 & 0 & 0 & 0 & 0 & 1 \\
\end{array}
\right),  
\hat{V} = 
\left(
\begin{array}{cccccccccccccccc}
 4 u & 1 & 0 & 0 & 1 & 0 & 0 & 0 & 0 & 0 & 0 & 0 & 0 & 0 & 0 & 0 \\
 1 & 0 & 1 & 0 & 0 & 1 & 0 & 0 & 0 & 0 & 0 & 0 & 0 & 0 & 0 & 0 \\
 0 & 1 & 0 & 1 & 0 & 0 & 1 & 0 & 0 & 0 & 0 & 0 & 0 & 0 & 0 & 0 \\
 0 & 0 & 1 & 0 & 0 & 0 & 0 & 1 & 0 & 0 & 0 & 0 & 0 & 0 & 0 & 0 \\
 1 & 0 & 0 & 0 & 0 & 1 & 0 & 0 & 1 & 0 & 0 & 0 & 0 & 0 & 0 & 0 \\
 0 & 1 & 0 & 0 & 1 & 4 u & 1 & 0 & 0 & 1 & 0 & 0 & 0 & 0 & 0 & 0 \\
 0 & 0 & 1 & 0 & 0 & 1 & 0 & 1 & 0 & 0 & 1 & 0 & 0 & 0 & 0 & 0 \\
 0 & 0 & 0 & 1 & 0 & 0 & 1 & 0 & 0 & 0 & 0 & 1 & 0 & 0 & 0 & 0 \\
 0 & 0 & 0 & 0 & 1 & 0 & 0 & 0 & 0 & 1 & 0 & 0 & 1 & 0 & 0 & 0 \\
 0 & 0 & 0 & 0 & 0 & 1 & 0 & 0 & 1 & 0 & 1 & 0 & 0 & 1 & 0 & 0 \\
 0 & 0 & 0 & 0 & 0 & 0 & 1 & 0 & 0 & 1 & 4 u & 1 & 0 & 0 & 1 & 0 \\
 0 & 0 & 0 & 0 & 0 & 0 & 0 & 1 & 0 & 0 & 1 & 0 & 0 & 0 & 0 & 1 \\
 0 & 0 & 0 & 0 & 0 & 0 & 0 & 0 & 1 & 0 & 0 & 0 & 0 & 1 & 0 & 0 \\
 0 & 0 & 0 & 0 & 0 & 0 & 0 & 0 & 0 & 1 & 0 & 0 & 1 & 0 & 1 & 0 \\
 0 & 0 & 0 & 0 & 0 & 0 & 0 & 0 & 0 & 0 & 1 & 0 & 0 & 1 & 0 & 1 \\
 0 & 0 & 0 & 0 & 0 & 0 & 0 & 0 & 0 & 0 & 0 & 1 & 0 & 0 & 1 & 4 u \\
\end{array}
\right).
\end{equation}
  At $\lambda =  0$, the ground state is again degenerate. Due to the size of the matrix, the analytical representation of the perturbation theory becomes cumbersome. It is thus preferable to study the ground state entanglement properties numerically. In Fig.~\ref{fig:F-H  Nup=Ndown=1 plot}, the numerical analysis of \eqref{app:eq:16 matrix} exhibits the power-law exponents as reported in the main text.

\section{Entanglement of a multiple conserved quantities: auxiliary proofs}

In the main text, the power-law decay of $\Delta S_{EE},\Delta S_q$ was obtained. The purpose of this section is to present the derivation in detail. 

First, we remind that representing $\ket{G^{(0)}},\ket{G^{(1)}}$ in the Schmidt decomposition as a superposition of  eigenstates of $\hat{N}^i _X$ was already demonstrated in appendix \ref{app:sec N1 aux proofs}.

Then, recall that for two conserved quantities, perturbation theory leads to the ground state written as 
\begin{eqnarray}
\ket{G^{(0)}} &=& \sum_{k,n_2} \alpha _{n_2,k} \ket{\Phi^k _{N_1,N_2-n_2} }_A \ket{\Psi^k _{0,n_2}}_B  \\  \nonumber 
\ket{G^{(1)}} &=& \sum_{n_1,n_2} \beta_{n_1,n_2,k} \ket{\phi^k _{N_1-n_1,N2_-n2}}_A \ket{\psi^k _{n_1,n2}}_B . 
\end{eqnarray}
So, it is left to present the calculation where the ground state form leads to the announced power-law decay, for both cases: I) Completely orthogonal sets $\Phi,\phi$ and $\Psi,\psi$ II) A single non-orthogonal pair $\Phi,\phi$.

\subsection{Entanglement decay with completely orthonormal sets }

Here we assume that 
\begin{equation}
    \braket{\Phi^k _{N_1,N_2-n_2}}{\phi^{k'} _{N_1-n_1,N_2-n' _2}} = \braket{\Psi^k _{0,n_2}}{\psi^{k'} _{n' _1,n' _2}} = 0 . 
\end{equation}

We can find $R$ as a function of $\lambda$
\begin{equation}
    R = \frac{N_1 \sum_{k,n_2} |\alpha_{k,n_2}|^2 + O(\lambda^2)}{\lambda^2 \sum_{k,n_1,n_2} n_1 |\beta_{k,n_1,n_2} |^2 }.  
\end{equation}
So, we find $R \propto N_1 / \lambda^2$. The normalized ground state to first order is $\ket{G} = \frac{1}{\sqrt{\mathcal{N}_\lambda}} (\ket{G^{(0)}}+\lambda \ket{G^{(1)}})$, where $\mathcal{N}_\lambda = \sum_{k,n_2} (\alpha_{k,n_1})^2 + \lambda^2 (\beta_{k,n_1,n_2})^2 = \mathcal{N}_0 + \lambda^2 \delta \mathcal{N} $. The Schmidt coefficients are the sets $ \frac{\alpha_{k,n_1}}{\sqrt{\mathcal{N}_\lambda}} $ and $\frac{\beta_{k,n_1,n_2}}{\sqrt{\mathcal{N}_\lambda}}$. From the above Schmidt coefficients, it is straight-forward to calculate the entanglement entropy. They both correspond to the announced results in the main text. Note that indeed $S_{EE}(\lambda \rightarrow 0) = - \sum_{k,n_1} \frac{|\alpha_{k,n_1}|^2}{\mathcal{N}_0} \ln \frac{|\alpha_{k,n_1}|^2}{\mathcal{N}_0}  \neq 0 $. Similarly, $S_q(\lambda \rightarrow 0)\neq 0 $.

\subsection{Entanglement decay with non-orthonormal sets}

Let us now consider all the pairs $\Psi,\psi$ to be orthonormal and all the pairs $\Phi,\phi$ orthonormal as well except for one non-orthogonal pair $\ket{v_1}   = \ket{\Phi^{k'} _{N_1,N_2-n' _2 }}$, $\ket{v_2} = \ket{\phi^{k'} _{N_1,N_2-n' _2 }}$ with their corresponding prefactors $\alpha_{n' _2,k'}, \beta_{0,n' _2 k' }$ being set to $a, b$ correspondingly. 

To leading order $R\propto N_1 / \lambda^2 $. This remains unchanged as in the previous case since $\langle \hat{N}^1 _A \rangle $ is still dominated by the $\lambda$ independent terms. Let us write the normalized ground state, traced out with respect to subsystem $B$: 
\begin{eqnarray}
    &\frac{1}{\mathcal{N}_\lambda}\sum_{ \Phi \neq v_1   } \alpha_{k,n_2} \ket{\Phi^k _{N_1,N_2-n_2}}
     +\frac{1}{\mathcal{N}_\lambda}\sum_{ \phi \neq v_2 } \alpha_{k,n_2} \ket{\phi^k _{N_1-n_1,N_2-n_2}}
    \\ \nonumber 
    &+ \frac{a}{\mathcal{N}_\lambda} \ket{v_1} + \frac{\lambda b}{\mathcal{N}_\lambda} \ket{v_2}.&  
\end{eqnarray}
Let us rewrite $a\ket{v_1}+\lambda b \ket{v_2}$ in an orthogonal fashion 
\begin{equation}
    a\ket{v_1} + \lambda b \ket{v_2} = (a+\lambda b \braket{v_1}{v_2})\ket{v_1} + \lambda b \sqrt{u}  \ket{v^\perp _2 }
\end{equation}
where $\ket{v^\perp _2} = u^{-1/2}(\hat{1}_v-\ket{v_1}\bra{v_1})\ket{v_2}$ and $u = 1- |\braket{v_1}{v_2}|^2$ and $\hat{1}_v$ is the unity operator in the subspace spanned by $v_{1,2}$. From the above, the normalized ground state, traced out with respect to subsystem $B$ can be represented as  
\begin{eqnarray}
    \frac{1}{\mathcal{N}_\lambda}\sum_{ \Phi \neq v_1   } \alpha_{k,n_2} \ket{\Phi^k _{N_1,N_2-n_2}}
    +\frac{1}{\mathcal{N}_\lambda}\sum_{ \phi \neq v_2 } \alpha_{k,n_2} \ket{\phi^k _{N_1-n_1,N_2-n_2}}
    \label{eq:F6}
    \\ \nonumber 
    + \frac{a+\lambda b \braket{v_1}{v_2}}{\mathcal{N}_\lambda} \ket{v_1} + \frac{\lambda b \sqrt{u} }{\mathcal{N}_\lambda} \ket{v^\perp _2}. 
\end{eqnarray}
Note that now the prefactors of the reduced ground state above (eq. F.6) 
are the Schmidt coefficients. We can calculate the excess entanglement entropy from the previous section, leading to another constant contribution (swallowed into $\Delta S_{EE}$) and another contribution scaling like $\lambda \propto 1/\sqrt{R}$. Notice that this contribution overtakes the $\ln R / R$ leading contribution of the orthogonal pairs. Namely, we infer that $\Delta S_{EE} \propto 1/\sqrt{R}$. Other non-orthogonal pairs, contributed additional $1/\sqrt{R}$ terms and hence do not change the universal character of the power-law. 
Similarly using eq. F.6, 
we find the results reported in the main text for $S_q$.

\end{document}